\documentclass{bmcart}

\usepackage{amsthm,amsmath}
\usepackage[utf8x]{inputenc}
\usepackage{grffile} 

\usepackage{newtxtext,newtxmath}
\usepackage{microtype}

\usepackage{amsmath}
\usepackage{amssymb}
\usepackage{bbm}
\usepackage{mathtools}
\usepackage{dsfont}
\usepackage{braket}
\usepackage{cancel}
\usepackage{slashed}
\usepackage{cite}

\usepackage{graphicx}

\usepackage{siunitx} 

\usepackage{ragged2e}
\usepackage{array}
\usepackage{tabularx}
\usepackage{booktabs}

\newcommand{\ii}{\operatorname{i}}
\newcommand{\expUp}[1]{\operatorname{e}^{#1}}
\newcommand{\dd}[1]{\mathrm{d} {#1} \;}

\renewcommand{\bra}[1]{\ensuremath{\langle #1|}}
\renewcommand{\ket}[1]{\ensuremath{|#1\rangle}}
\renewcommand {\braket}[2]{\ensuremath{\langle #1|#2\rangle}}

\startlocaldefs
\endlocaldefs
\definecolor{cset-aps-blueberry}{RGB}{28,128,158}
\definecolor{cset-aps-blue}{RGB}{46,44,184}
\definecolor{cset-aps-turquoise}{RGB}{0,67,88}
\definecolor{cset-aps-limegreen}{RGB}{190,219,67}
\definecolor{cset-aps-green}{RGB}{31,138,112}
\definecolor{cset-aps-yellow}{RGB}{255,225,25}
\definecolor{cset-aps-orange}{RGB}{253,116,0}
\definecolor{cset-aps-red}{RGB}{219,0,43}
\usepackage{hyperref}
\hypersetup{%
    colorlinks=true,
    linkcolor={cset-aps-red},
    linkbordercolor={cset-aps-red},
    filecolor={cset-aps-orange},
    filebordercolor={cset-aps-orange},
    citecolor={cset-aps-blue},
    citebordercolor={cset-aps-blue},
    urlcolor={cset-aps-green},
    urlbordercolor={cset-aps-green},
    menucolor={cset-aps-limegreen},
    menubordercolor={cset-aps-limegreen},
    breaklinks=true,
    pdfborderstyle={/S/U/W 2},
    pdfpagemode=UseOutlines,
    pdfstartpage={1},
}

\usepackage{tikz}

\usepackage{pgfplots}
\pgfplotsset{%
    every axis legend/.append style={%
        cells={anchor=west},
        at={(0.96,0.04)},
        anchor=south east,
        font=\scriptsize,
        },
    every axis/.append style={%
        yticklabel style={%
            /pgf/number format/fixed zerofill,
            /pgf/number format/precision=2},
        },
    width= \textwidth,
    height=8cm,
    xmajorgrids=true,
    xminorgrids=false,
    minor x tick num=1,
}

\usetikzlibrary{decorations}

\begin{document}

\begin{frontmatter}

\begin{fmbox}
\dochead{Research Article}


\title{Tunneling Gravimetry}


\author[
  addressref={aff1},                   
  email={patrik.schach@tu-darmstadt.de \\
  patrik.schach@gmail.com}   
]{\inits{P.}\fnm{Patrik} \snm{Schach}}
\author[
  addressref={aff2},
  email={alexander.friedrich@uni-ulm.de}
]{\inits{A. }\fnm{Alexander} \snm{Friedrich}}
\author[
  addressref={aff4},
  email={jason.r.williams.dr@jpl.nasa.gov}
]{\inits{J.R.}\fnm{Jason R.}  \snm{Williams}}
\author[
  addressref={aff2,aff3},
  email={wolfgang.schleich@uni-ulm.de}
]{\inits{W.P.}\fnm{Wolfgang P.}  \snm{Schleich}}
\author[
  addressref={aff1},
  email={enno.giese@tu-darmstadt.de}
]{\inits{E.}\fnm{Enno} \snm{Giese}}


\address[id=aff1]{%
    \orgdiv{Technische Universit{\"a}t Darmstadt, Fachbereich Physik, Institut f{\"u}r Angewandte Physik},
  \street{Schlossgartenstr. 7},
  \postcode{D-64289 Darmstadt},
  \cny{Germany}                                    
}
\address[id=aff2]{%
  \orgdiv{Institut f{\"u}r Quantenphysik and Center for Integrated Quantum Science and Technology (IQ\textsuperscript{ST})},
  \orgname{Universit{\"a}t Ulm},
  \street{Albert-Einstein-Allee 11},
  \postcode{D-89069 Ulm},
  \cny{Germany}
}
\address[id=aff3]{%
  \orgdiv{Hagler Institute for Advanced Study and Department of Physics and Astronomy, Institute for Quantum Science and Engineering (IQSE)},
  \orgname{Texas A{\&}M University},
  \postcode{TX 77843-4242},
  \city{College Station},
  \cny{USA}
}

\address[id=aff4]{%
  \orgdiv{Jet Propulsion Laboratory},
  \orgname{California Institute of Technology},
  \city{Pasadena},
  \postcode{CA 91109},
  \cny{USA}
}


\begin{artnotes}
\note{This article has been published in \href{https://doi.org/10.1140/epjqt/s40507-022-00140-3}{EPJ Quantum Technology \textbf{9}, 20 (2022)} under the terms of the \href{https://creativecommons.org/licenses/by/4.0/}{Creative Commons Attribution 4.0 International} license.}     
\end{artnotes}

\end{fmbox}


\begin{abstractbox}

\begin{abstract} 
We examine the prospects of utilizing matter-wave Fabry--P\'{e}rot interferometers for enhanced inertial sensing applications.
Our study explores such tunneling-based sensors for the measurement of accelerations in two configurations:
(a) a transmission setup, where the initial wave packet is transmitted through the cavity and (b) an out-tunneling scheme with intra-cavity generated initial states lacking a classical counterpart.
We perform numerical simulations of the complete dynamics of the quantum wave packet, investigate the tunneling through a matter-wave cavity formed by realistic optical potentials and determine the impact of interactions between atoms.
As a consequence we estimate the prospective sensitivities to inertial forces for both proposed configurations and show their feasibility for serving as inertial sensors.

\end{abstract}




\end{abstractbox}
%

\end{frontmatter}



\section{Introduction}

Matter-wave sensors allow for high-precision measurements of inertial forces.
Promising candidates include light-pulse atom interferometers~\cite{Kasevich1991, Cronin2009, VarennaLectures2014, Muntinga2013}, where beam splitters are realized by diffraction from optical lattices~\cite{Hartmann2020}, and guided interferometers~\cite{Ryu2015, Akatsuka2017, Navez2016} building on atomtronic circuits~\cite{Amico2021, Pandey2021}. 
Both approaches are based on the interference of matter waves propagating along two spatially separated paths. 
In comparison, the interference effects inside a cavity that generates a matter-wave Fabry--P\'{e}rot interferometer (FPI)~\cite{Wilkens1993, Carusotto2001, Valagiannopoulos2019, Dutt2010, Ruschhaupt2005, Manju2020, Manju_Thesis2020} cause distinct tunneling resonances. 
In this article, we propose two configurations of a matter-wave FPI and study their suitability for accelerometry or gravimetry. 

Optical Fabry--P\'{e}rot interferometers~\cite{Ismail16, Rajibul2014} consist of two mirrors that form a cavity.
Depending on the properties of the cavity and the wavelength of the incident light, interference effects cause peaks in the transmission spectrum.
These resonances can be used to filter a specific wavelength of the incident light beam so that the FPI acts as a monochromator~\cite{Geake1959}.   
As a consequence, FPIs can be used as accelerometers~\cite{Rajibul2014}:
An accelerated or falling cavity only transmits Doppler-shifted resonance frequencies.
In this case, the (massive) mirrors are accelerated (e.g., in a gravitational field), but the effects on the light waves with their vanishing rest mass are suppressed by the inverse speed of light $1/c$, as highlighted by the gravitational redshift as a first-order effect in $1/c^2$.
In contrast, matter waves couple strongly to gravity and have developed into a versatile tool for gravimetry and accelerometry~\cite{Peters1999, McGuirk2002, Wu2019, Stray2022}.

In order to generalize the concept of optical FPIs to matter-wave experiments~\cite{Wilkens1993, Carusotto2001, Valagiannopoulos2019, Dutt2010, Ruschhaupt2005, Manju2020, Manju_Thesis2020}, we replace the mirrors by optical potentials that act as a barrier for the incident matter wave.
Such barriers can be generated~\cite{Bergamini2004, Zupancic2016} or painted~\cite{Henderson2009} experimentally with the help of spatial light modulators~\cite{McGloin2003, Boyer2006}, digital (micro)mirror devices~\cite{Gauthier2016, Zupancic2016}, and acousto-optic deflectors~\cite{Trypogeorgos2013, Bell2016}. 
Tunneling~\cite{Carr2005, Manju2018,Lindberg2021} leads to a transmission of the incident matter wave though the barrier. 
It is a purely quantum effect without any counterpart in classical mechanics that is used to conventionally describe massive particles.
While optical interferometers can be used as monochromators, matter-wave FPIs can act as velocity~\cite{Ruschhaupt2005} and angular filters~\cite{Valagiannopoulos2019} for matter waves. 

One prominent application of such interferometers in the field of atomtronics is the use of nonlinear interactions between ultra-cold atoms to obtain a similar behavior to the Coulomb blockade in tunnel junctions~\cite{Carusotto2001} or to create highly entangled many-particle states~\cite{Haug2019}.
Repulsive self-interaction and finite momentum width of the atomic cloud lead to a suppression of the resonances of the matter-wave cavity~\cite{Manju2020, Manju_Thesis2020}.
To our knowledge, Reference~\cite{Manju_Thesis2020} constitutes the only prior research in this context and has broached the possibility of a matter-wave FPI as an acceleration sensor, based on analytical models.

In this article, we explore the feasibility of developing matter-wave FPIs for practical accelerometry. 
In contrast to the previous study, we introduce realistic cavities that are distorted by gravity and use a numerical time evolution to obtain the uncertainty of acceleration measurements.
We introduce experimental observables to identify and isolate two physical effects susceptible to gravity:
(i) velocity filtering of the matter-wave cavity and (ii) distortion of the transmission spectrum itself. 
By that, we highlight the potential of matter-wave FPIs for applications in inertial sensing and bridge the gap between light-pulse atom interferometry and atomtronics.

In Sec.~\ref{sec:TransSpec} we study the scattering of a matter wave incident on a cavity made of two optical barriers and discuss the influence of gravity on the transmission spectrum as well as the sensitivity of such a setup to gravity. 
The main contribution is acceleration of the wave packet prior to scattering.
To highlight the effect of gravity on the matter-wave cavity itself, we determine in Sec.~\ref{sec:CavityResonances} its resonances and their respective widths. 
In addition, we discuss the deformation of the wave packet scattered from gravity-distorted barriers.
In Sec.~\ref{sec:AsymTunneling}, we study a situation where the wave packet is prepared inside the cavity in a superposition of counter-propagating momenta.
In contrast to an incident wave packet, where the acceleration prior to the interaction dominates, such a configuration isolates the effect of gravity on the cavity itself. 
As a consequence the transmission through the left and right barriers differs and can be used as a sensing device for gravitational or inertial forces.
We conclude with a summary and future prospects of this quantum sensor in Sec.~\ref{sec:Discussion}.

\section{Transmission Spectroscopy}
\label{sec:TransSpec}

\subsection{Setup}

\begin{figure}[h!]
	\centering
    \includegraphics{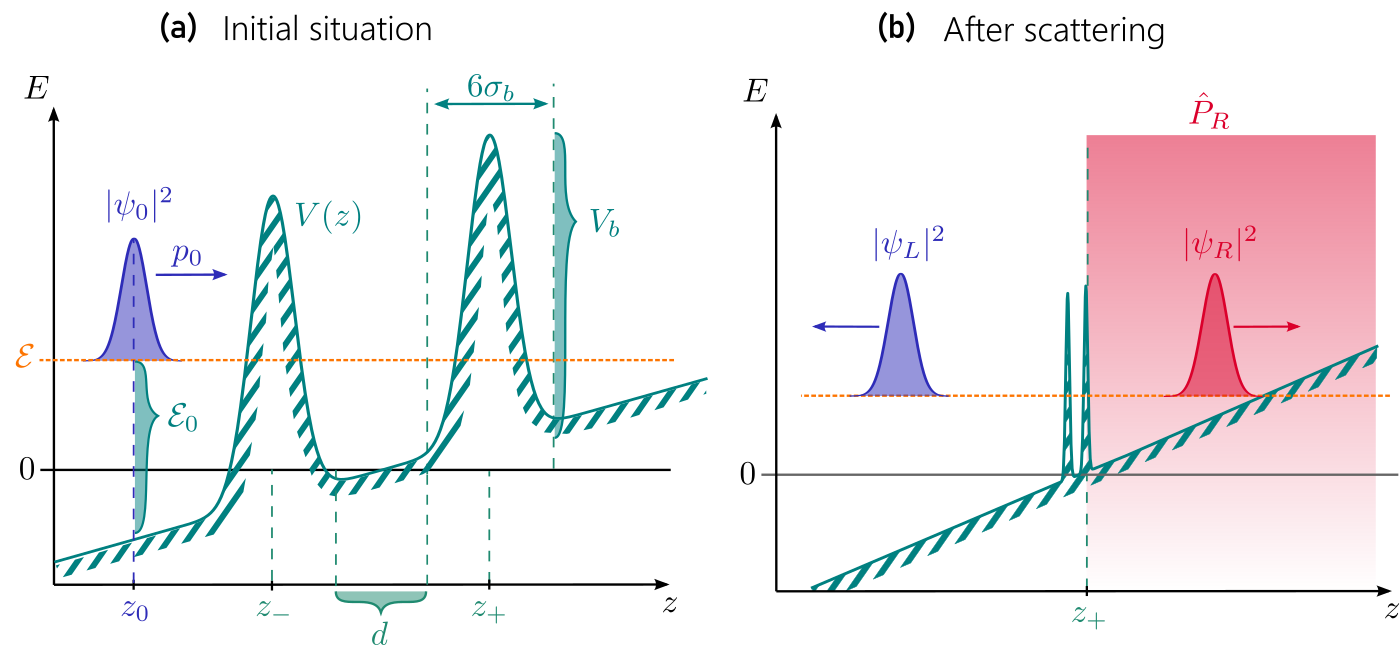}
    \caption{
    Gravitationally distorted matter-wave cavity and wave packet (a) prior to scattering and (b) after scattering.
    The matter-wave cavity consists of two Gaussian barriers with height $V_b$ and width $\sigma_b$, at positions $z = z_\pm$, chosen such that the overlap between both barriers is negligible for vanishing cavity length $d$.
    (a) The initial wave packet $\psi_0$ is located at $z_0$ and has an initial momentum $p_0$ that corresponds to the kinetic energy $\mathcal{E}_0 = p_0^2 / (2 m)$.
    The gravitational field disturbs the propagation of the wave packet and the matter-wave cavity.
    To account for the influence of the gravitational field $g$, we take the kinetic energy $\mathcal{E} = \mathcal{E}_0 - m g |z_0|$ at the center of the matter-wave cavity as reference. 
    (b) The initial wave packet scatters from the matter-wave cavity, resulting in a superposition of reflected and transmitted wave packet $|\psi_L|^2$ and $|\psi_R|^2$.
    To obtain the number of transmitted atoms, we introduce the operator $\hat{P}_R^2 = \hat{P}_R$ that projects on the region to the right of the cavity (shaded in red).
    }
    \label{fig:ConfigStartOutside}
\end{figure}

Figure~\ref{fig:ConfigStartOutside}(a) shows the matter-wave Fabry--P\'{e}rot cavity consisting of two Gaussian barriers of height $V_b$ and width $\sigma_b$ located at $z_\pm = \pm (3 \sigma_b + d/2)$, such that for a vanishing cavity length $d=0$, the overlap between both barriers is negligible.
Gaussian-shaped barriers represent a realistic approximation of optical potentials that can be generated~\cite{Bergamini2004, Zupancic2016} or painted~\cite{Henderson2009} experimentally.
We consider the case where the initial wave packet is tightly confined in the transverse direction within a quasi one-dimensional waveguide, where the transverse length scale is smaller than the Rayleigh range of the light beams generating the Gaussian barriers. 
Perturbations of the cold atomic cloud associated with the transverse confinement are therefore neglected in our treatment. 
When the matter-wave cavity is disturbed by a gravitational field leading to the acceleration $g$, the Gaussian barriers become asymmetric and the motion of the wave packet is affected.
The transmission of a guided wave packet through the gravitationally disturbed matter-wave cavity is modeled by the time-dependent one-dimensional Schrödinger equation
\begin{equation}
    \ii \hbar \frac{\mathrm{d}}{\mathrm{d} t} \ket{\psi} = \left[ \frac{\hat{p}^2}{2 m} + m g \hat{z} + V_b \left\{ \exp{\left( - \frac{(\hat{z} - z_-)^2}{2 \sigma_b^2} \right)} + \exp{\left( - \frac{(\hat{z} - z_+)^2}{2 \sigma_b^2} \right)} \right\} \right] \ket{\psi} ,
    \label{eq:SETotalSystem}
\end{equation}
where $\hbar$ denotes the reduced Planck constant and $m$ the mass of the atoms described by the wave packet $\braket{z}{\psi}$ at time $t$.
The position and momentum operators $\hat{z}$ and $\hat{p}$ fulfill the commutation relation $[\hat{z}, \hat{p}] = \ii \hbar$, where the corresponding position and momentum eigenstates are defined by $\hat{z}\ket{z}=z\ket{z}$ and $\hat{p}\ket{p}=p\ket{p}$.

For the initial state, we assume a Gaussian wave packet with position variance $\Delta z$ and position $z_0 = - 3 \Delta z - 6 \sigma_b - d/2$ defined via
\begin{equation}
    \psi_0(z) = \braket{z}{\psi_0} = \frac{1}{(2 \pi \Delta z^2)^{1/4}} \exp{\left( - \frac{(z - z_0)^2}{4 \Delta z^2} + \ii \frac{(z - z_0) p_0}{\hbar} \right)}
    \label{eq:initial_wave_packet}
\end{equation}
in position representation. 
As a consequence, the overlap between the initial wave packet and the matter-wave cavity is exponentially small and thus negligible.
The initial momentum $p_0$ of the wave packet can be imparted, e.g., via Bragg or Raman diffraction \cite{Giese2015} and the small momentum width $\Delta p = \hbar /(2 \Delta z)$ required for the interferometer can be prepared via Delta-kick collimation \cite{Ammann1997, Muntinga2013}.
Figure~\ref{fig:ConfigStartOutside}(b) shows the scattered wave packet represented as superposition of reflected and transmitted wave packets that depend on the properties of the matter-wave cavity as well as gravitational acceleration.

\subsection{Transmission spectra}

To describe the fraction of transmitted atoms, we introduce the operator $\hat{P}_R^2 = \hat{P}_R$ that projects on the space to the right of the matter-wave cavity, shown in Fig.~\ref{fig:ConfigStartOutside}(b).
Using the projector $\hat{P}_R$, the fraction of transmitted atoms becomes \cite{ReedSimon1979}
\begin{equation}
    T_R = \bra{\psi_{sc}} \hat{P}_R \ket{\psi_{sc}}
    = \int\limits_{z_+}^\infty \dd{z} |\psi_{sc}(z)|^2
    = \int\limits_{-\infty}^\infty \dd{p} |\tau(p)|^2 |\psi_0(p)|^2
    \label{eq:AveragingFormula}
\end{equation}
where $\psi_{sc}(z) = \braket{z}{\psi_{sc}}$ describes the scattered wave function after a finite time $t$, which is long enough so that no population is observed inside the cavity.
The last equality only holds in the limit $t \to \infty$ for vanishing gravitational acceleration $g=0$. 
If the wave packet is launched against gravity, i.e., for  $g > 0$, the transmitted wave packet will eventually impinge a second time on the cavity after passing the apex of its trajectory.
We exclude such bouncing effects by choosing an appropriate finite time in our simulations, where any population that remains in the cavity can be neglected ($<1\%$) and the transmitted wave packet has not yet returned.
In position representation the projector reduces to an integral of the scattered wave function from the position $z_+$ of the second barrier to infinity.
For the description in momentum representation without gravity we consider the decomposition of the initial wave packet into momentum eigenstates $p$ with individual transmission amplitudes $\tau(p)$ that give rise to the transmission spectrum $|\tau(p)|^2$.
Thus, the total transmission coefficient $T_R$ of a wave packet reduces to the transmission coefficients for individual momentum eigenstates weighted by the initial momentum distribution $|\psi_0(p)|^2$.
Both representations coincide in the asymptotic limit $t \to \infty$ while the numerical simulation ends after a sufficiently large but finite time. 
The details of the implementation are discussed below.  

\begin{figure}[h!]
	\centering
    \includegraphics[scale=1.3]{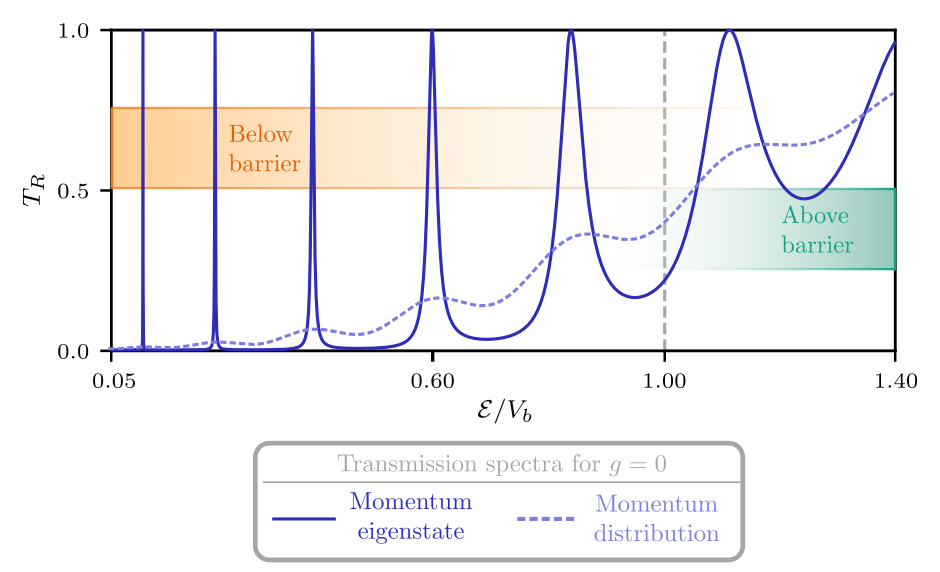}
    \caption{
    Transmission spectrum of a matter-wave cavity for a momentum eigenstate (solid blue) with kinetic energy $\mathcal{E}$ and for a momentum distribution (dashed blue, $\Delta z = \SI{12}{\text{\textmu}\meter}$) as well as vanishing gravitational acceleration $g = 0$. 
    The peaks in the transmission spectrum give rise to resonances whose widths increase for larger kinetic energies. 
    We scale the distance between two resonances by the cavity length (here $d=\SI{15}{\text{\textmu}\meter}$) and the width of the individual resonances is related to the width of the barriers (here $\sigma_b = \SI{1}{\text{\textmu}\meter}$, $V_b=\SI{1.42e-25}{\joule}$).
    The spectrum is obtained by the transfer matrix approach and discretizing the total transfer matrix into $10^2$ sub-matrices.
    }
    \label{fig:TransSpectrum}
\end{figure}

\subsubsection{Plane waves}
The transmission spectrum in Fig.~\ref{fig:TransSpectrum} for vanishing accelerations is obtained via the transfer matrix ansatz following Refs. \cite{Jirauschek2009, Loran2020} that allows for the determination of transmission coefficients for momentum eigenstates.
The transfer matrix relates the wave function to the left of the cavity with the wave function to the right of the matter-wave cavity.
Consequently, its elements contain information about the transmission process. 
The Gaussian barriers are approximated by step-wise potentials, each described by a transfer matrix that can be determined analytically.
The total transfer matrix is given by concatenation of the substeps, as a consequence of the semi-group property of transfer matrices~\cite{Loran2020}.
Increasing the accuracy of the approximation, the number of transfer matrices increases and , in the limit of infinite number, the approach becomes exact.
For our simulation, we increased the number of steps so that the result converged.
Our set of parameters gave rise to a step size of $\SI{0.27}{\text{\textmu}\meter}$.
An additional possibility to obtain the transmission spectrum of the matter-wave cavity is given by the WKB approximation~\cite{Ankerhold2007,Dutt2010} that uses a semi-classical expansion to obtain an approximate wave function.
However, this semi-classical technique is only valid close to the top of the barrier and is therefore not suited to treat the tunneling of the narrow resonances at low energies, which are at the focus of this article. 
We therefore refrain from presenting a comparison.

Calculating the fraction of transmitted atoms for different momenta corresponding to the momentum eigenstates $p$ leads to the transmission spectrum depicted in Fig.~\ref{fig:TransSpectrum}.
The distinct peaks in the spectrum give rise to resonances that depend on the properties of the matter-wave cavity. 
Compared to the optical FPI, the reflectivity of the mirrors corresponds to the barrier width and the distance between the mirrors to the distance separating the Gaussian barriers. 
In the transmission spectrum we observe sharp resonances for low kinetic energies and broader peaks for larger energies, leading to an overlap of several resonances at larger kinetic energies. 
Resonances corresponding to larger energies are bound more weakly due to their vicinity to the continuum and the decreasing width of the barriers, resulting in shorter lifetimes and subsequently broader resonances. 
In analogy to the optical cavity, the matter-wave cavity filters specific momenta of the initial wave packet, and acts as a monochromator for sufficiently narrow resonances. 

Approximating a gravitationally distorted matter-wave cavity by rectangular barriers with asymmetric heights allows to define the transmission of a momentum eigenstate analytically similar to an optical FPI~\cite{Manju_Thesis2020}.
With the help of Eq.~\eqref{eq:AveragingFormula}, these results can be used to find the transmission of a broad wave packet.
Finite momentum widths and interatomic interaction of atomic clouds suppress the resonances, conversely reducing the interaction and the momentum width improves the resonant peaks.

\subsubsection{Wave packets}

We determine the numerical time-evolution of a wave packet scattered from a matter-wave cavity consisting of two realistic barriers by a Fourier-split step method.
The upper limit of integration in position as well as the lower and upper limit in momentum coincide with the end of the position and momentum grids.
Moreover, the final time of the numerical evolution is chosen such that a negligible fraction of atoms is left inside the cavity ($<1 \% $) and the overlap of reflected and transmitted waves with the matter-wave cavity is negligible.

The gravitational field disturbs the motion of the wave packet and consequently influences the fraction of transmitted atoms. 
To give greater insight into the resonances of the matter-wave cavity, we study the effect of the gravitational acceleration $g$ on the wave packet close to microgravity~\cite{Muntinga2013} as well as the influence of the initial momentum $p_0$.
As reference scale, we chose the kinetic energy $\mathcal{E}$ at the center of the matter-wave cavity and as a consequence acceleration and deceleration of the wave packet, depending on the sign of the gravitational acceleration, are included in the initial kinetic energy $\mathcal{E}_0 = \mathcal{E} + m g |z_0|$.
In particular, this choice sets a lower bound to the initial momentum for negative gravitational accelerations.

\begin{figure}[h!]
	\centering
    \includegraphics{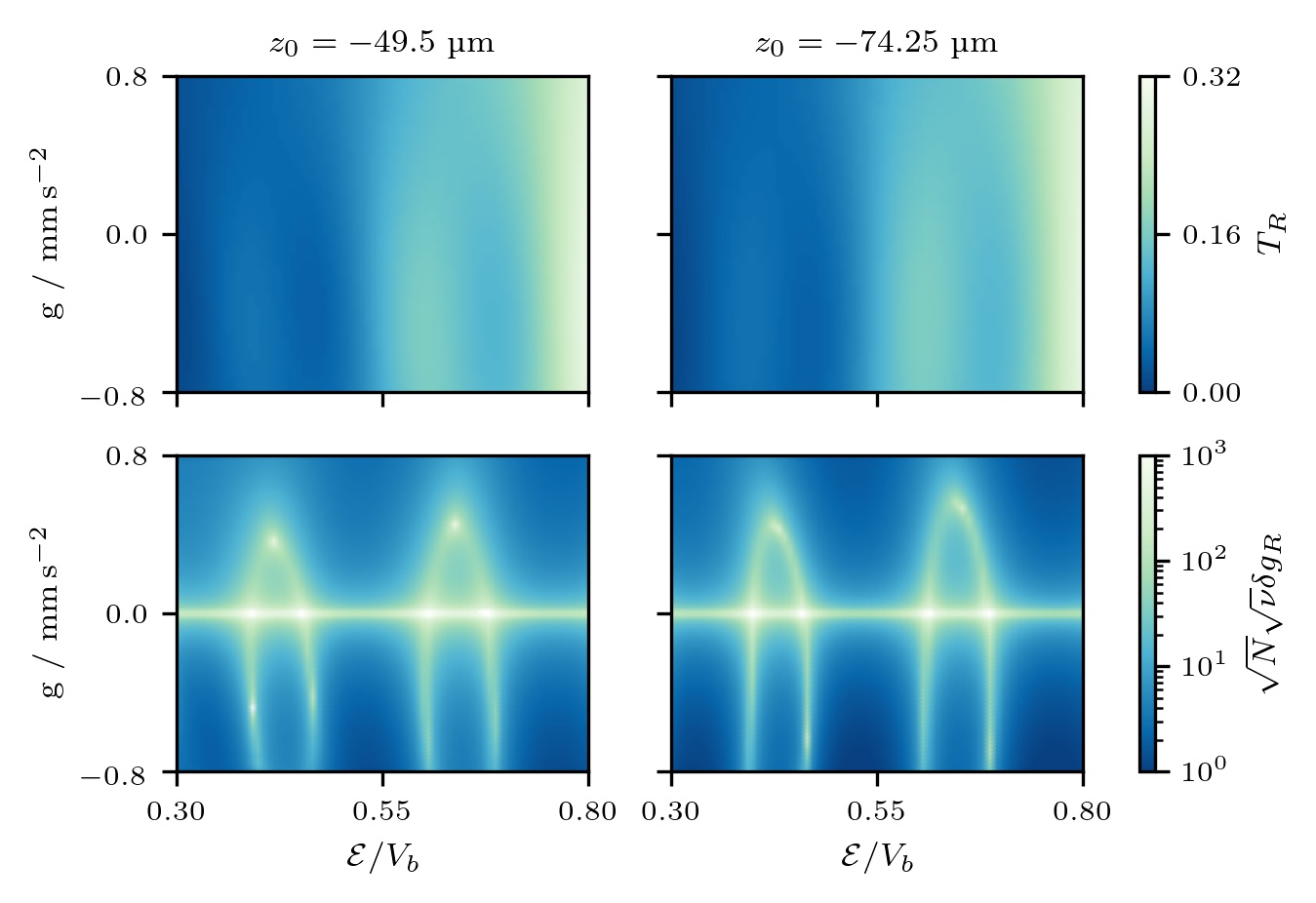}
    \caption{  
    Transmission (top) of two wave packets with different initial positions $z_0$ (left and right) under the influence of gravity and the relative uncertainty (bottom) of these gravimeters.
    Initially, the wave packet (initial width $\Delta z = \SI{12}{\text{\textmu}\meter}$) receives the momentum kick $p_0$ and subsequently scatters from the matter-wave cavity ($\sigma_b = \SI{1}{\text{\textmu}\meter}$, $V_b=\SI{1.42e-25}{\joule}$, and $d=\SI{15}{\text{\textmu}\meter}$).
    We chose the final time of numerical evolution $t_f=\SI{1}{\second}$ to ensure a negligible fraction of atoms remain inside the cavity. 
    To take into account the influence of the gravitational field $g$ prior to scattering, we take the kinetic energy $\mathcal{E} = \mathcal{E}_0 - m g |z_0|$ as reference where $\mathcal{E}_0 = p_0^2 / (2 m)$ describes the initial kinetic energy and $z_0$ the initial position of the wave packet. 
    The resonances in transmission (top) occur for the same momenta as for momentum eigenstates, but are less prominent due to the finite width $\Delta z$ of the wave packet. 
    A similar effect is induced by gravity, so that the resonances wash out for $g > 0$, while they are more prominent for $g < 0$.
    The relative uncertainty (bottom) estimates the sensitivity of the matter-wave cavity with respect to gravity by a measurement of the fraction of transmitted atoms. 
    For small gravitational accelerations the relative uncertainty diverges.
    This effect is represented by white, visualizing relative uncertainties that exceed the maximum value of the colorbar.
    The regions of minimal uncertainty (dark blue) define the desired working points of the sensor.
    While $\delta g_R$ denotes the relative uncertainty for an experiment with $N$ particles and $\nu$ repetitions, we plot the quantity $\sqrt{N}\sqrt{\nu}\delta g_R$ which is the single-particle uncertainty without repetitions, assuming shot-noise limited measurements with non-interacting particles.
    }
    \label{fig:TransSpecWavePacket_z0}
\end{figure}

The transmission spectrum for a wave packet, in a gravitational field, scattered from the matter-wave cavity is shown in Fig.~\ref{fig:TransSpecWavePacket_z0}.
We observe no significant shift of the resonances due to the gravitational field, but for $g > 0$ the resonances wash out, in contrast to $g < 0$ where the resonances become more prominent. 
Due to gravity, the barriers of the matter-wave cavity are asymmetric and thus the wave packet experiences a different slope while coupling into the cavity. 
Consequently, the momentum width of the wave packet gets distorted depending on the value of $g$ and modifies the transmission.
In the next section we study the effects of wave packet deformations in more detail. 
The matter-wave cavity still acts as a monochromator, that is, only a specific class of momenta is transmitted and therefore the propagation prior to the interaction is severely affected. 
In summary, the mean kinetic energy and the width of the wave packet are crucial when coupling into the matter wave cavity.

\subsection{Sensitivity to accelerations}

To quantify the uncertainty of an acceleration measurement, we use Gaussian error propagation~\cite{Taylor1997} and find for the error of $T_R=T_R(g,\mathcal{E})$ the relation
\begin{equation}
    \Delta T_R^2 =  | \partial_g T_R |^2 \Delta g_R ^2 +  | \partial_\mathcal{E} T_R |^2 \Delta \mathcal{E}^2 ,
\end{equation}
where $\Delta g_R$ is the uncertainty of the gravitational acceleration obtained from a transmission measurement and $\Delta \mathcal{E}$ the error of $\mathcal{E}=\mathcal{E}_0 - m g |z_0|$.
If we assume that the initial position and initial kinetic energy are known with certainty, we can connect $\Delta \mathcal{E}=  | m z_0| \Delta g_R$ to the uncertainty of the acceleration and hence obtain 
\begin{equation}
    \delta g_R = \frac{\Delta g_R}{\sqrt{N} \sqrt{\nu} g} = \frac{\Delta T_R}{\sqrt{N} \sqrt{\nu}  \sqrt{| g \partial_g T_R |^2 + |m g z_0 \partial_\mathcal{E} T_R  |^2} } .
    \label{eq:RelUncR}
\end{equation}
Here, we have included the number of measurements $\nu$ as well as the number of atoms $N$ in the atomic cloud.
If we use the variance of the observable as a measure for the uncertainty of the transmission, i.e., $\Delta T_R^2 = \bra{\psi_{sc}} \hat{P}_R^2 \ket{\psi_{sc}}-\bra{\psi_{sc}} \hat{P}_R \ket{\psi_{sc}}^2$, and use the idempotence of the projector $\hat{P}_R^2=\hat{P}_R$, we obtain the relative uncertainty $\delta g_R$ directly from the transmission spectrum with the help of Eq.~\eqref{eq:RelUncR} and $\Delta T_R^2 = T_R (1-T_R)$.

Another measure beyond Gaussian error propagation is the classical Fisher information~\cite{Giovannetti2011} that gives rise to the sensitivity obtained for a specific measurement and observable.
For pure states and analytical expressions, the Fisher information can be computed straightforwardly.
However, there is no analytical solution for the quantum state after tunneling through gravitationally distorted Gaussian barriers.
One approach to treat the problem in an analytical manner~\cite{Manju_Thesis2020} is to approximate the cavity by two perfectly rectangular barriers of different heights.
This height difference is chosen to correspond to the potential difference caused by gravity over the cavity length so that the linear potential inside the cavity can be neglected.
In this way one can obtain an expression for the classical Fisher information of momentum eigenstates.
One can also include the effects of wave packets by averaging over such eigenstates, in analogy to the semi-analytical model introduced above.
In this case, optimizing the cavity length depending on the initial wave packet leads to an increased sensitivity.
The overall sensitivities in such a simple model are of the same order of magnitude as those observed below.
Moreover, we will show that the width of the wave packet is affected by realistic Gaussian barriers so that the results of the optimization procedure cannot be easily transferred to our setup. 

Inspection of Eq.~\eqref{eq:RelUncR} shows that two different effects contribute to the relative uncertainty:
(i) The dependence of the transmission spectrum on accelerations itself is included in $| g \partial_g T_R |$. 
(ii) The contribution $| m g z_0 \partial_\mathcal{E} T_R |$ is caused by the propagation prior to impact and scales with the initial position $z_0$.

For one experimental run, the cold atomic cloud of width $\Delta z$ is prepared at position $z_0$ and experiences a momentum kick $p_0$, e.g., imparted by Bragg diffraction~\cite{Giese2015}.
Subsequently, the atoms propagate in the gravitational field causing a shift of the momentum distribution, as expected from a drop experiment. 
When the atoms impinge on the barriers, the filtering properties of the resonances therefore provide a measure for the shift of the momentum distribution.
In this sense, the experiment can be seen as a conventional drop experiment that measures a momentum distribution in the near field.
However, the resonances in the transmission spectrum themselves are also distorted by gravity, an effect that has no analogy in a drop experiment.

\subsubsection{Estimates and comparison to light-pulse atom interferometers}
The sensitivity of the matter-wave FPI obtained from Eq.~\eqref{eq:RelUncR} is shown in Fig.~\ref{fig:TransSpecWavePacket_z0} and reaches sensitivities up to $\sqrt{N} \sqrt{\nu} \delta g_R \approx 2$ for a single particle and single run.
To compare the sensitivity of the FPI to Mach-Zehnder atom interferometers (MZI), we assume a preparation time of approximately $\SI{300}{ms}$ and $10^7$ atoms of rubidium 87~\cite{Menoret2018}.
Close to the optimal working point at $\mathcal{E} / V_b= 0.77 $ and $g = \SI{-0.8}{mm s^{-2}}$ for $z_0=\SI{-49.5}{\text{\textmu}\meter}$, the sensitivity of the FPI is $\SI{500}{nm \, s^{-2} \, Hz^{-1/2}}$ with the corresponding duration of the experiment of $\SI{350}{ms}$.
In current experiments~\cite{Menoret2018} with MZIs, sensitivities of approximately $\SI{500}{nm \, s^{-2} \, Hz^{-1/2}}$ with a repetition rate of $\SI{2}{Hz}$ have been achieved.
For comparison, the sensitivity of classical sensors based on MEMS~\cite{Mustafazade2020, Tang2019} are around $\SI{1}{\text{\textmu} m \, s^{-2} \, Hz^{-1/2}}$ depending on the experiential design. 
The estimated sensitivities of the FPI and the experimentally achieved sensitivities of the MZI are of the same order.
However, we have omitted any uncertainty of preparation of the initial wave packet that limits the sensitivity in our analysis and compared a theoretical analysis to an actual experiment.
In fact, the initial conditions have to be verified by separate measurements akin to drop experiments.
Hence, the estimates underline the potential of the presented technique, highlighting only its intrinsic limitations without claiming a competitiveness to an atomic MZI.

While the sensitivity of the MZI to accelerations scales quadratically with interferometer time, the required free-fall distance of the matter waves scales the same way.
Some terrestrial atom interferometer experiments now operate with matter waves spanning on the order of tens of meters~\cite{Schilling2020}.
In contrast, the interaction region of the FPI considered in our study is relatively small, namely around $\SI{30}{\text{\textmu}\meter}$.
In fact, in Ref.~\cite{Manju_Thesis2020} the size of the interaction region was used as a figure of merit for a comparison of both concepts, but this is only one factor.
For example, comparing experiments of the same duration can still lead to a different result.

Since the beam splitters and mirrors of a standard MZI are implemented by diffraction of the matter waves from counterpropagating light fields, where the laser phase difference is imprinted onto the diffracted component, such devices are sensitive to laser phase noise.
On the other hand a matter-wave FPI measures the fraction of transmitted atoms, and since the optical potentials depend on the intensity of the optical field, they are insensitive to laser phases.
Therefore, the matter-wave FPI is a robust sensor with respect to laser-phase instabilities associated with the optical barriers and is operated in a compact interaction region that is favorable for miniaturized quantum sensors~\cite{Abend2016}.

\subsubsection{Connection to Bragg spectroscopy}
We have demonstrated that the matter-wave cavity acts as a monochromator and that this velocity filtering has a major impact on its sensitivity to accelerations.
The effect can be seen as a measurement of the momentum distribution in the near field, after a certain time of acceleration.
To compare the monochromator properties of the matter-wave FPI to Bragg spectroscopy~\cite{Stenger1999, Papp2008, Veeravalli2008}, we consider the same experimental setup as before:
A cold atomic cloud of width $\Delta z$ at position $z_0$ is released in a gravitational field with initial momentum $p_0$.
After a certain time of free fall, we apply long velocity-selective Bragg pulses with an effective Rabi frequency $\Omega$ to diffract a fraction of the expanded wave packet with a momentum transfer $\hbar k_B$.
For our set of parameters and a gravitational acceleration of $0.5 \, \mathrm{mm \, s^{-2}}$ a time of free fall of about $60 \, \mathrm{ms}$ corresponds to the distance used for transmission spectroscopy.
Because the diffraction process is sensitive to the Doppler detuning, it can also be used to determine the momentum distribution after some acceleration.
For a mirror pulse neglecting higher-order diffraction, we define the velocity width through a Doppler detuning $\nu_\text{FW}$, which has the dimensionless form $\varepsilon_\text{FW} = \nu_\text{FW}/\Omega= 1.597 $ and is determined by $1/2 = (\pi/2)^2 \operatorname{sinc}^2(\pi/2 \sqrt{1 + \varepsilon_\text{FW}})$, i.e., by the full width at half maximum.
We find that the velocity selectivity of the Bragg pulse is comparable to that of a matter-wave cavity if the condition $\hbar \Omega_j = \Gamma_j \hbar k_B/(2\sqrt{2 m E_{r,j}}\varepsilon_{FW})$ is satisfied.
Here, $E_{r,j}$ is the energy of a particular resonance and $\Gamma_j$ its width.
The numerical procedure to find these resonances is discussed in Sec.~\ref{sec:CavityResonances}. 
For resonances comparable to the matter-wave cavity considered in this article we obtain the Bragg Rabi frequencies shown in Tab.~\ref{tab:rabi_frequencies}. 
If compared to typical Bragg pulses, which have Rabi frequencies on the order of kilohertz, the frequencies shown in Tab.~\ref{tab:rabi_frequencies} correspond to extremely long pulse durations.
Moreover, such a measurement scheme also suffers from the uncertainty of the initial conditions. 
\begin{table}
    \centering
    \caption{Resonance energies $E_{r,j}$, and widths $\Gamma_j$ of a matter-wave FPI which gives the same velocity selectivity of a Bragg pulse with effective Rabi frequency $\Omega_j$ (barrier width $\sigma_b = \SI{1}{\text{\textmu}\meter}$, barrier height $V_b=\SI{1.42e-25}{\joule}$, dimensionless Doppler width $\epsilon_\text{FW}=1.597$, mass $m=\SI{1.4431609e-25}{\kilogram}$ of $^{87}\mathrm{Rb}$ and cavity length $d=\SI{15}{\text{\textmu}\meter}$).}
    \begin{tabular}{lrr}
         \toprule
         $E_{r,j}/V_b$ & $\Gamma_j/V_b$ & $\Omega_j/(2\pi) $ \\
         \midrule
         $0.03$ & $2.23 \times 10^{-5}$ &  $0.04\,\mathrm{Hz}$ \\
         $0.10$ & $2.61 \times 10^{-4}$ &  $0.24\,\mathrm{Hz}$ \\
         $0.23$ & $1.44 \times 10^{-3}$ &  $0.89\,\mathrm{Hz}$ \\
         $0.39$ & $5.22 \times 10^{-3}$ &  $2.45\,\mathrm{Hz}$ \\
         $0.60$ & $0.02 $ &  $6.46\,\mathrm{Hz}$ \\
         $0.83$ & $0.05 $ & $15.45\,\mathrm{Hz}$ \\
         $1.11$ & $0.11$  & $31.99\,\mathrm{Hz}$ \\
         \bottomrule
    \end{tabular}
    \label{tab:rabi_frequencies}
\end{table}

\begin{figure}[h!]
	\centering
    \includegraphics{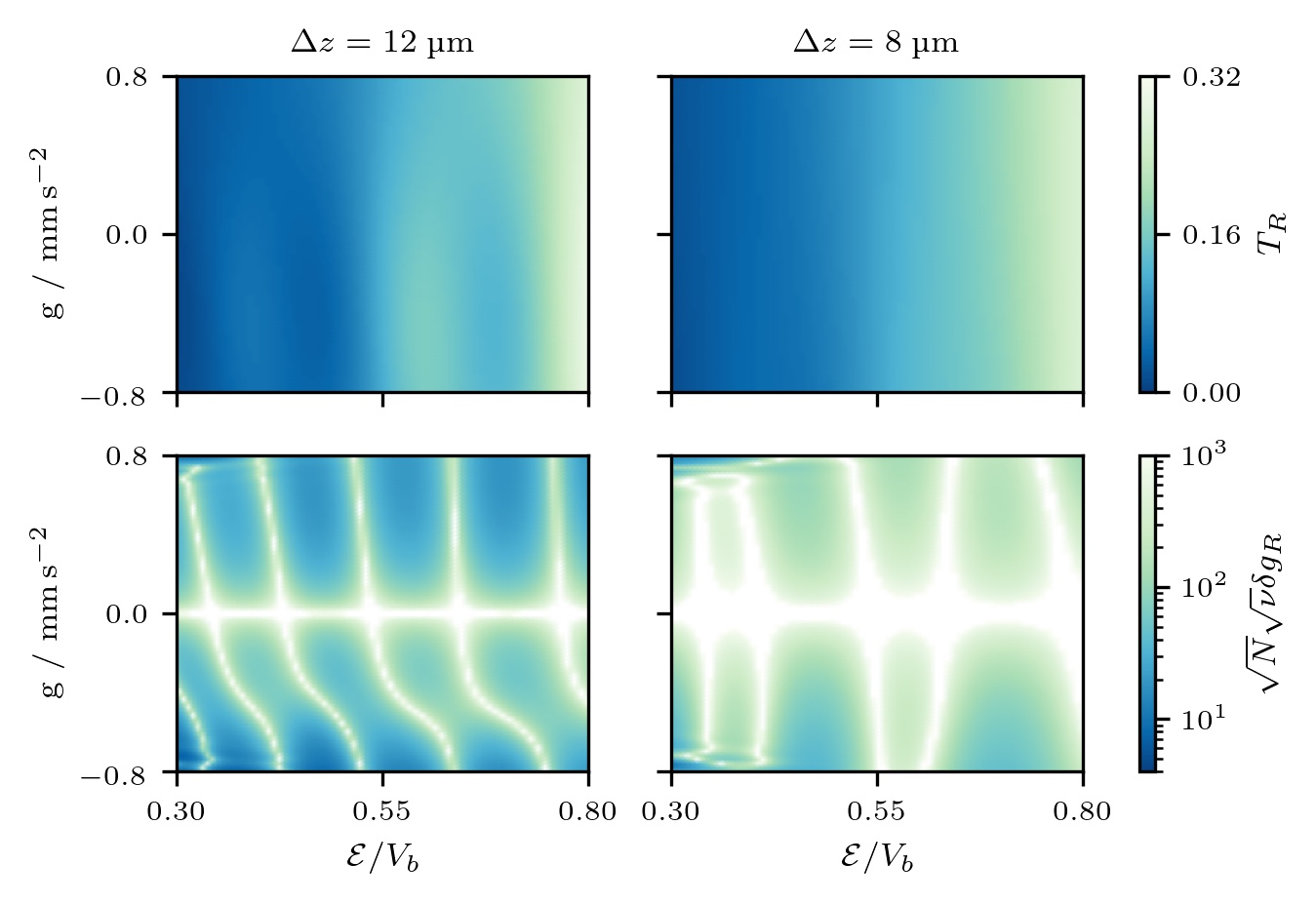}
    \caption{
    Transmission (top) of two wave packets of different initial widths $\Delta z$ (left and right) under the influence of gravity and the relative uncertainty (bottom) of such a gravimeter.
    Initially, the wave packet receives a momentum kick $p_0$ and subsequently scatters from the matter-wave cavity ($\sigma_b = \SI{1}{\text{\textmu}\meter}$, $V_b=\SI{1.42e-25}{\joule}$, and $d=\SI{15}{\text{\textmu}\meter}$).
    We chose the final time of numerical evolution $t_f=\SI{1}{\second}$ to ensure a negligible fraction of atoms remain inside the cavity. 
    To take into account the influence of the gravitational field $g$ prior to scattering, we used the kinetic energy $\mathcal{E} = \mathcal{E}_0 - m g |z_0|$ at the center of the cavity as a reference, where $\mathcal{E}_0 = p_0^2 / (2 m)$ describes the initial kinetic energy and $z_0$ the initial position of the wave packet.
    The resonances in the transmission (top) occur for the same momenta as for momentum eigenstates, but are less prominent due to the finite width $\Delta z$ of the wave packet. 
    A similar effect is induced by gravity, so that the resonances wash out for $g > 0$, while they are more prominent for $g < 0$.
    The relative uncertainty (bottom) estimates the sensitivity of the matter-wave cavity with respect to gravity by a measurement of the fraction of transmitted atoms. 
    Here, we omitted the term $| m g z_0 \partial_\mathcal{E} T_R |$ in Eq.~\eqref{eq:RelUncR} to isolate the effect of the matter-wave cavity.
    For small gravitational accelerations the relative uncertainty diverges.
    This effect is represented by white, visualizing relative uncertainties that exceed the maximum value of the colorbar.
    The regions of minimal uncertainty (dark blue) define the desired working points of the sensor.
    While $\delta g_R$ denotes the relative uncertainty for an experiment with $N$ particles and $\nu$ repetitions, we plot the quantity $\sqrt{N}\sqrt{\nu}\delta g_R$ which is the single-particle uncertainty without repetitions, assuming shot-noise limited measurements with non-interacting particles.   
    }
    \label{fig:TransSpecWavePacket}
\end{figure}

\subsubsection{Exclusion of propagation effects}
We have omitted so far any errors of the initial conditions, which severely deteriorate the sensitivity of the sensor.
The matter-wave cavity filters the momentum distribution at impact and thus the influence of the gravitational acceleration on the wave packet prior to scattering can be determined. 
However, the initial position and by that the duration of the acceleration period needs to be known with sufficient precision, in analogy to drop experiments.
To estimate the intrinsic sensitivity of the matter-wave cavity, we omit the contribution $| m g z_0 \partial_\mathcal{E} T_R |$ to the denominator in Eq.~\eqref{eq:RelUncR}.
The obtained sensitivities together with the transmission spectra of two wave packets with different initial widths are shown in Fig.~\ref{fig:TransSpecWavePacket}. 
The position of the resonances is independent of the wave packet's width, but the resonances are more prominent for narrower initial momentum distributions.
The sensitivity of the matter-wave FPI contains regions of minimal uncertainty with an optimum $\sqrt{N} \sqrt{\nu} \delta g_R \approx 6$ defining possible working points of a sensor.
Here, the best sensitivity is $\SI{1.8}{\text{\textmu} m \, s^{-2} \, Hz^{-1/2}}$ at the working point $\mathcal{E}/V_b = 0.3 $ and $g = \SI{-0.8}{mm \, s^{-2}}$ for $\Delta z=\SI{12}{\text{\textmu}\meter}$ as well as the duration of the experiment of $\SI{700}{ms}$, a preparation time of approximately $\SI{300}{ms}$, and $10^7$ atoms of rubidium 87.
As expected, omitting the contribution that stems from the propagation and scales with the initial condition decreases sensitivity by approximately one order of magnitude. 

Even though these estimates are purely academic, they highlight that the properties of the matter-wave cavity itself are susceptible to accelerations beyond the filtering effect that can be interpreted as an analogue of a drop experiment.  
We therefore study in the next section the distortions of the matter-wave cavity before we devise a setup in Sec.~\ref{sec:AsymTunneling} that isolates this effect. 

\section{Distortions induced by gravity}
\label{sec:CavityResonances}

In the previous section we have seen that the acceleration of the wave packet prior to the interaction with the matter-wave cavity contributes significantly to the sensitivity of the sensor.
However, the distortion of the cavity by gravity gives rise to an additional contribution.
Before we propose a setup to isolate the effect in Sec.~\ref{sec:AsymTunneling}, we first shed light on the resonances of the cavity under the influence of gravity.
Although the asymptotics of the matter-wave cavity permits no bound states independent of the presence of a gravitational potential, resonances manifest themselves in quasi-bound states~\cite{Carr2005, Uma2009} that have finite lifetimes and thus finite decay widths. 
In general, the Hamiltonian $\hat{H}(\hat{z}, \hat{p})$ is hermitian and subsequently possesses a real spectrum. 
To find finite lifetimes of quasi-bound states, we introduce a phase $\theta$ and apply the complex scaling~\cite{Baye2015}
\begin{equation}
    \quad \hat{z} \to \hat{z} \expUp{\ii \theta} \quad \hat{p} \to \hat{p} \expUp{-\ii \theta} .
\end{equation}
As a consequence, the Hamiltonian $\hat{H}(\hat{z} \expUp{\ii \theta}, \hat{p} \expUp{-\ii \theta})$ ceases to be hermitian and possesses complex eigenvalues $E_j = E_{r,j} - \ii \Gamma_j / 2$ as well as non-orthogonal eigenstates.  
We identify $E_{r,j}$ with the energy and $\Gamma_j$ with the width associated with the resonances, in addition $\Gamma_j^{-1}$ denotes the lifetime of the corresponding quasi-bound state. 
To describe quasi-bound states, we select specific solutions of the time-independent Schrödinger equation in position representation
\begin{equation}
    \left[ -\frac{\hbar^2}{2 m} \frac{\partial^2}{\partial z^2} \expUp{-\ii \theta} + V\left(z \expUp{\ii \theta} \right) \right] \psi_{\varphi_j}(z) = \left( E_{r,j} - \ii \Gamma_j / 2 \right) \psi_{\varphi_j}(z) .
    \label{eq:HamiltonianComplexScaling}
\end{equation}
Moreover, we impose boundary conditions to obtain states with exponentially decaying tails, independent of the gravitational background potential. 
For that, we assume
\begin{equation}
    \psi_{\varphi_j}(z) =
    \begin{cases}
		a_L \expUp{- \ii k_j z \expUp{\ii \theta}}, & z \to -\infty \\
        a_R \expUp{\ii k_j z \expUp{\ii \theta}}, & z \to +\infty
	\end{cases}
	\label{eq:BoundaryCondResStates}
\end{equation}
with amplitudes $a_L$ and $a_R$ as well as the free wave vector $\hbar k_j = \sqrt{2m E_j}$.
To find an explicit condition for the phase $\theta$, we express the free wave vector $\hbar k_j = \sqrt{2m E_j} = \sqrt{2m |E_j|} \expUp{\ii \varphi_j / 2}$ with the complex eigenvalues $E_j = |E_j| \expUp{-\ii \varphi_j}$ where $\varphi_j = \operatorname{arctan}(\Gamma_j / (2 E_{r,j}))$ denotes the angle enclosed with the real axis.
We find that for $2 \theta > \varphi_j$ the states $\psi_{\varphi_j}(z)$ satisfy vanishing boundary conditions, subsequently the states describe quasi-bound states with finite lifetimes.

\begin{figure}[h!]
	\centering
    \includegraphics{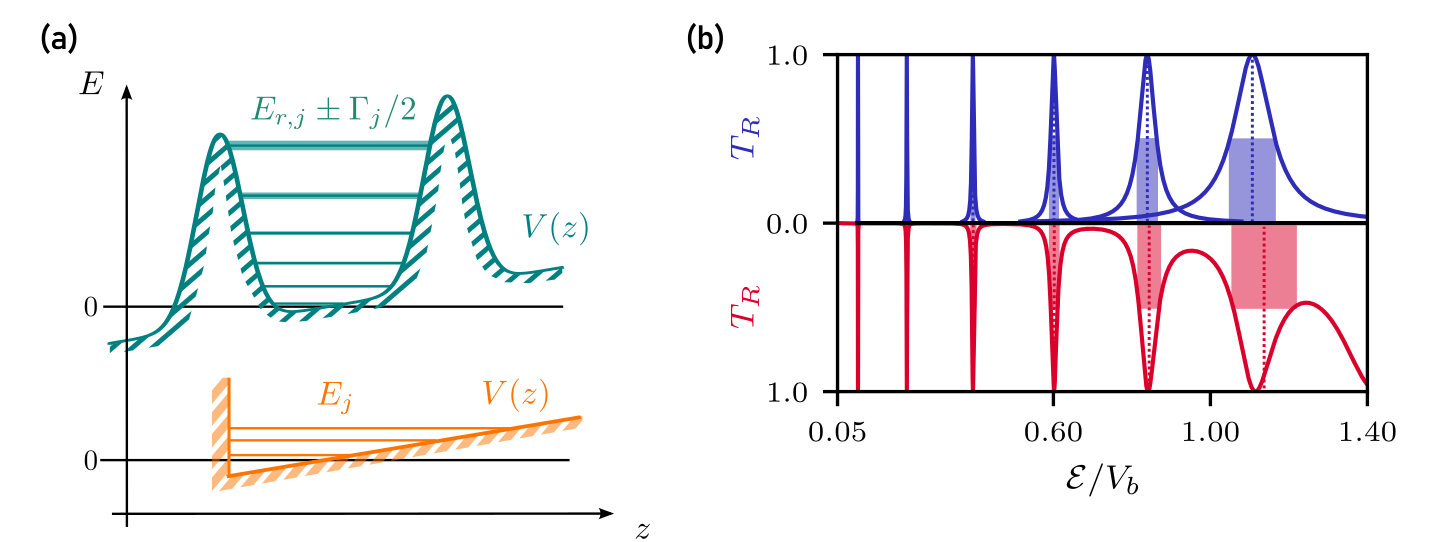}
    \caption{
    (a) Resonances and bound states of the distorted matter-wave cavity (top) and a triangular potential (bottom). 
    The resonances (green solid lines) of the matter-wave cavity (green) are modeled by quasi-bound states whose lifetime correspond to the width (shaded line) of the respective resonance.
    In contrast, the asymptotic of the triangular potential (orange) allows bound states (orange solid lines).
    To compare both cases, the left boundary of the triangular potential (orange) is chosen such that it correspond to the center of the left barrier of the matter-wave cavity (green).
    For a strong acceleration, we expect the lower resonances of the matter-wave cavity to approach the eigenenergies of the triangular potential.
    (b) Comparison between the transmission spectrum (red solid line) from Fig. \ref{fig:TransSpectrum} and Lorentzian profiles (blue solid lines), defined in Eq.~\eqref{eq:ResonanceLorentzian}, associated with the individual resonances for $g=0$.
    The eigenenergies and widths are obtained by a Lagrange-mesh method which diagonalizes the Hamiltonian that describes the matter-wave cavity ($\sigma_b = \SI{1}{\text{\textmu}\meter}$, $V_b=\SI{1.42e-25}{\joule}$, and $d=\SI{15}{\text{\textmu}\meter}$).
    The Lorentzian profiles show good agreement with the transmission spectrum while the overlap between resonances is negligible.
    }
    \label{fig:ResWidthsLorentz}
\end{figure}

Figure~\ref{fig:ResWidthsLorentz}(a) shows the eigenvalues of the non-hermitian Hamiltonian obtained via the Lagrange-mesh method~\cite{Baye2015} which diagonalizes the Hamiltonian that describes the matter-wave cavity.
To compare the obtained resonances with the transmission spectrum in Fig.~\ref{fig:TransSpectrum}, we assume that the shape of a individual resonance $j$ is given by a Lorentzian~\cite{Uma2009}
\begin{equation}
    f_j(\mathcal{E}) = \frac{(\Gamma_j/2)^2}{(\mathcal{E} - E_{r,j})^2 + (\Gamma_j/2)^2}
    \label{eq:ResonanceLorentzian}
\end{equation}
where full width half maximum (FWHM) of the Lorentzian corresponds to the width of the resonance $\Gamma_j$ and the expectation value to the energy $E_{r,j}$ associated with the resonance.
Comparing the Lorentzian profiles to the transmission spectrum, as shown in Fig~\ref{fig:ResWidthsLorentz}(b), we find good agreement for the expectation values and widths of the Lorentzian profiles.
This simple method does not consider the overlap of several resonances but the relevant features of the lower resonances are captured well. 

\begin{figure}[h!]
	\centering
    \includegraphics[width=0.9\textwidth]{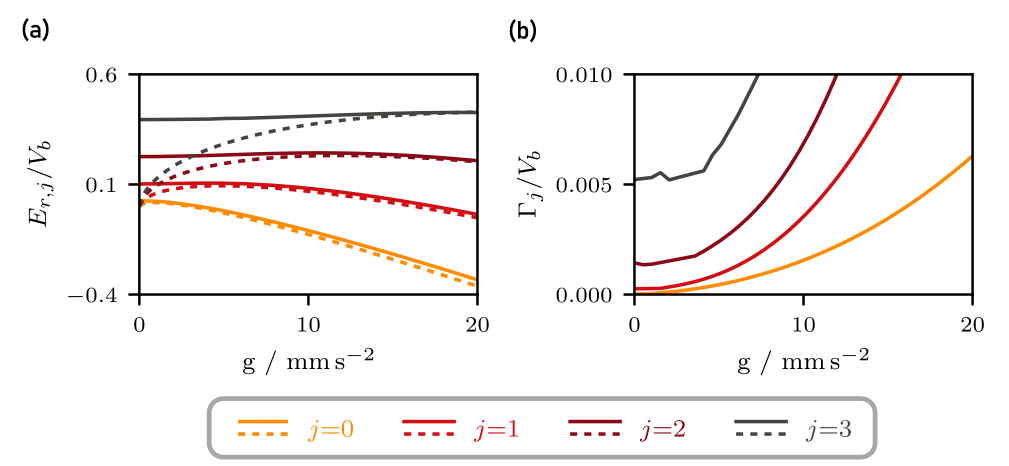}
    \caption{
    Influence of gravity on the resonances (a) and their widths (b) of a matter wave cavity ($\sigma_b = \SI{1}{\text{\textmu}\meter}$, $V_b=\SI{1.42e-25}{\joule}$ and $d=\SI{15}{\text{\textmu}\meter}$).
    The plotted values correspond to the real and imaginary parts of the eigenvalues of the complex-scaled Hamiltonian, obtained by the Lagrange-mesh method.
    In addition to the resonances (solid lines), the eigenenergies of the triangular potential shown in Fig.~\ref{fig:ResWidthsLorentz}(a) are included (dashed lines). 
    If the gravitational acceleration is sufficiently strong, we expect that the right barrier becomes less important and bound states arise solely from the left barrier and the linear potential. 
    As a consequence, the energies associated with the resonances approach the eigenenergies of the triangular potential (dashed lines). 
    Moreover, for larger gravitational accelerations the resonances are closer to the continuum resulting in shorter lifetimes and subsequently larger widths of the resonances.
    }
    \label{fig:GravResonances}
\end{figure}

The influence of accelerations on the resonances and their widths is shown in Fig.~\ref{fig:GravResonances} together with bound state energies of the triangular potential sketched in Fig.~\ref{fig:ResWidthsLorentz}(a). 
The gravitational acceleration induces an asymmetry between both barriers and modifies the energies of the resonances.
Already a single barrier in a gravitational potential leads to (gravitationally) bound states.
Hence, if the variation of the gravitational potential is of the order of the barrier height over the length of the cavity, the effect of the second barrier becomes irrelevant.
As a consequence, we expect the resonance energies to approach the eigenenergies of the ideal triangular potential for large accelerations.
Indeed, we observe this effect in Fig.~\ref{fig:GravResonances}.
Moreover, for larger accelerations the width of all resonances increases, even though the effect is more dominant for resonances that correspond to larger energies.
This behavior can be understood the following way:
The width of a resonance increases if it is closer to the continuum of unbound states.
Moreover, for higher energies the states are not as strongly bound, since the observed Graussian barrier becomes smaller.
Therefore the lifetime decreases, which in turn leads to an increased width of the resonance. 
The gravitational acceleration effectively changes the height of the barriers and introduces an asymmetry so that all resonances are closer to the continuum of unbound states.

\begin{figure}[h!]
	\centering
    \includegraphics[width=0.9\textwidth]{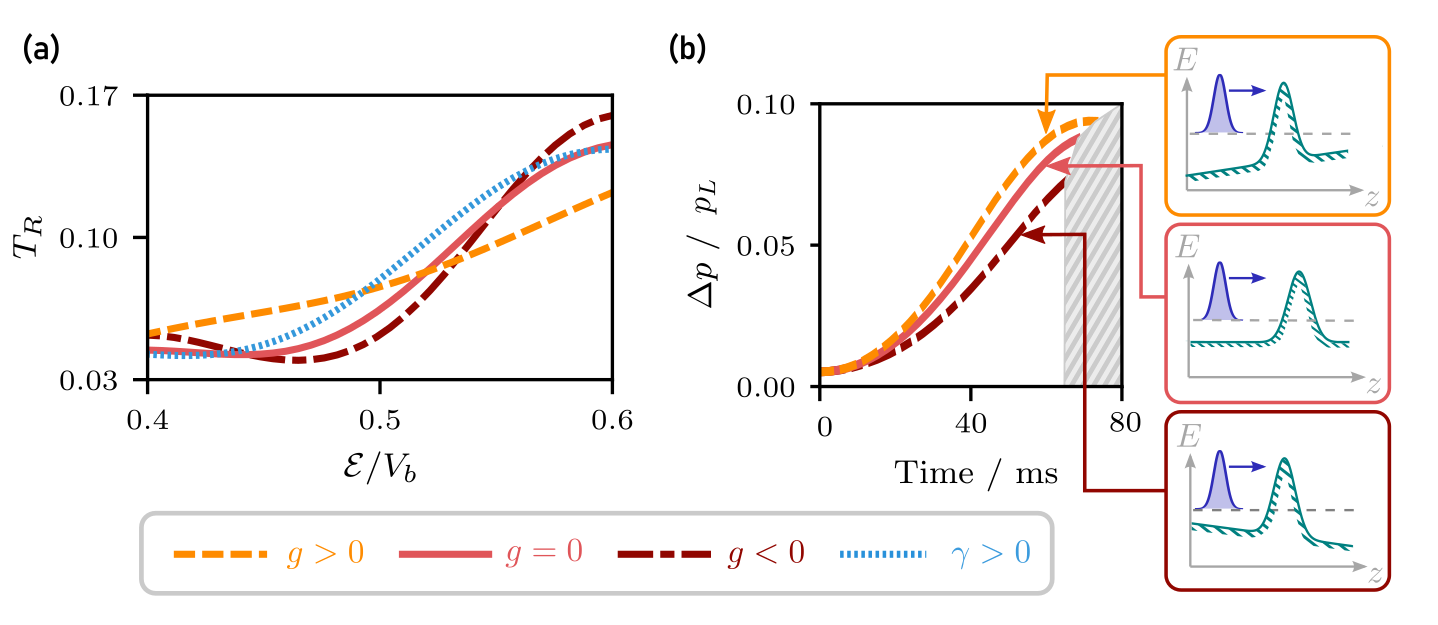}
    \caption{
    (a) Transmission of a wave packet (initial width $\Delta z = \SI{12}{\text{\textmu}\meter}$) with initial momentum $p_0$ scattered from the matter-wave cavity ($\sigma_b = \SI{1}{\text{\textmu}\meter}$, $V_b=\SI{1.42e-25}{\joule}$, $d=\SI{15}{\text{\textmu}\meter}$).
    To take into account the influence of the gravitational field $g$ prior to scattering, we use the kinetic energy $\mathcal{E} = \mathcal{E}_0 - m g |z_0|$ as reference where $\mathcal{E}_0$ describes the initial kinetic energy and $z_0$ the initial position of the wave packet.
    Without considering the self-interaction of the atomic cloud, the resonances wash out for $g > 0$ ($g = \SI{1.3}{mm/s^2}$) and become more prominent for $g < 0$ ($g = \SI{-0.8}{mm/s^2}$). 
    A repulsive self-interaction $\gamma > 0$ (here $\gamma = \SI{3.51e-38}{m}$, $g = \SI{0}{mm/s^2}$) leads to a suppression of the resonances (dashed line).
    (b) Momentum width of the time-evolved wave packet.
    The individual plots end at the time of the turning point of a classical particle with same momentum $\mathcal{E}/V_b=0.77 $. 
    The momentum width $\Delta p$ is scaled by $p_L = m v_R$ with the recoil velocity $v_R = \SI{5.8845}{mm/s}$ of the $^{87}\mathrm{Rb}$ $\mathrm{D_2}$-transition.
    The slope of the the barriers is affected by gravity and in turn deforms the wave packet upon propagation.
    The effect of the direction of gravity is shown in the insets to the right.
    As a consequence, the wave packet contracts in momentum for $g < 0$, while the width is increased for $g > 0$.
    }
    \label{fig:TransWidths}
\end{figure}

Although we observe resonances in the transmission spectra, shown in Fig.~\ref{fig:TransWidths}(a), mainly the width of the resonances varies and the corresponding energy is not shifted.
That is, the structure of the transmission spectrum washes out for $g > 0$ and becomes more concentrated for $g < 0$.
In addition, large negative accelerations lead to initial kinetic energies larger than the Gaussian barriers and consequently no resonances are observable. We note that the approximately quadratic asymptotic dependence of the resonance widths $\Gamma_j$ as a function of gravitational acceleration $g$ is not surprising and can be understood as a direct consequence of Fermi's golden rule applied to the case of a fixed energy state decaying into the surrounding continuum \cite{Cohen2019}. 

To gain more insight into the asymmetry between positive and negative accelerations, we study the momentum width of the wave packet.
The widths of the momentum distribution of the time-evolved wave packet, up to the impact on the matter-wave cavity, is shown in Fig.~\ref{fig:TransWidths}(b).
For $g > 0$ the momentum width is larger and for $g < 0$ smaller than in the case of no gravitational acceleration, attributed to different slopes of the asymmetrically distorted barriers.
Since, during the coupling into the cavity, the momentum width determines the resolution of the transmission spectrum of a wave packet, the resonances wash out for $g > 0$ and become prominent for $g < 0$.

A similar effect can be observed for a self-interacting quantum gas like a Bose--Einstein condensate.
For that, we describe the time-evolution by the one-dimensional Gross-Pitaevskii equation 
\begin{equation}
    \ii \hbar \frac{\partial \psi(z)}{\partial t}
    = \left[- \frac{\hbar^2}{2m} \frac{\partial^2}{\partial z^2} + V(z) + \gamma |\psi(z)|^2 \right] \psi(z)
\end{equation}
using the mean-field approximation, where $\gamma$ is the strength of the self-interaction and $V(z)$ described the potential used in Eq.~\eqref{eq:SETotalSystem}.
Figure~\ref{fig:TransWidths}(a) includes the fraction of transmitted atoms of a wave packet governed by the Gross-Pitaevskii equation.
Considering repulsive self-interaction $\gamma > 0$ during propagation, the self-energy of the quantum gas converts to kinetic energy, resulting in an increase of the wave packet's momentum width.
Consequently, the structure of the transmission spectrum washes out, as observed in Figure~\ref{fig:TransWidths}(a). 
The reverse effect can be observed for sufficiently small attractive self-interaction if its spatial extend is larger than the soliton size~\cite{Khaykovich2002}.

\section{Asymmetric Tunneling}
\label{sec:AsymTunneling}

\begin{figure}[h!]
	\centering
    \includegraphics{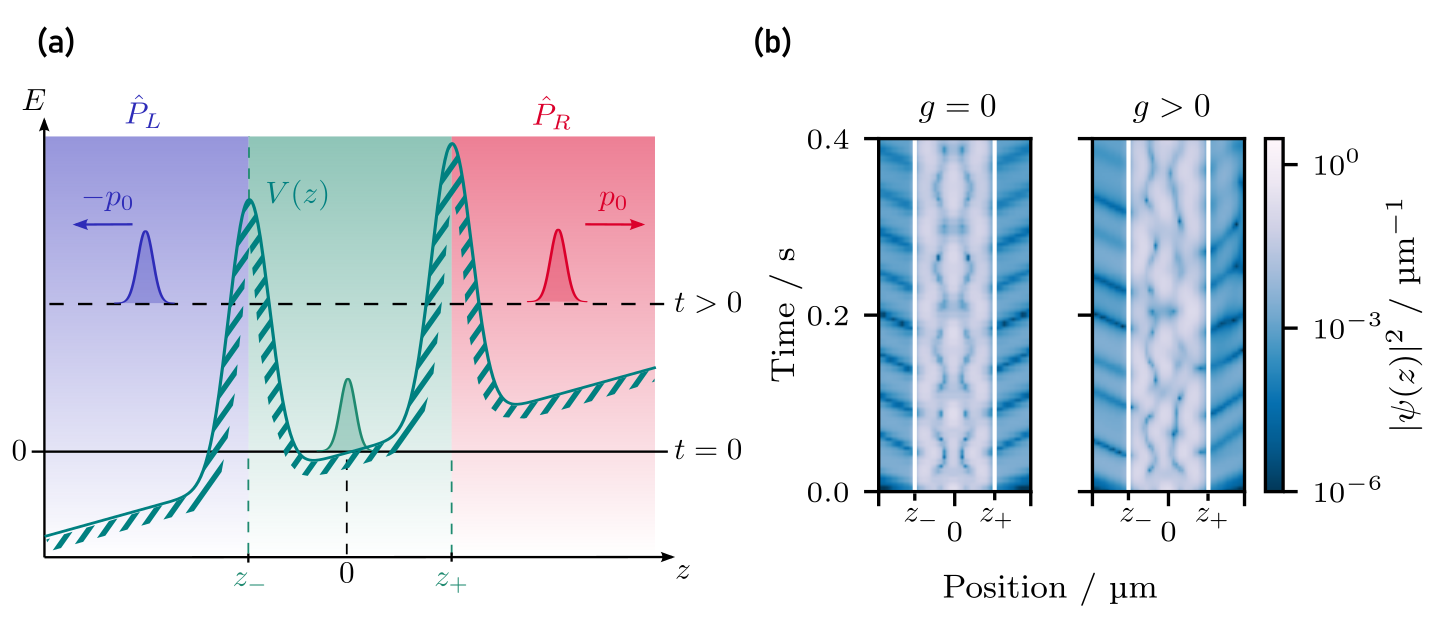}
    \caption{
    (a) Gaussian wave packet ($\Delta z = \SI{3}{\text{\textmu}\meter}$) prepared in the center of the gravitationally distorted matter-wave cavity ($\sigma_b = \SI{1}{\text{\textmu}\meter}$, $V_b=\SI{1.42e-25}{\joule}$, $d=\SI{15}{\text{\textmu}\meter}$). 
    A double Bragg pulse creates a superposition of the wave packet with opposite momenta $\pm p_0$.
    Due to gravity, the transmission of the kicked wave packet through the left (blue region) and right barrier (red region) differs.
    The operators $\hat{P}_L$ and $\hat{P}_R$ project on the fraction of transmitted atoms through the left (blue shaded region) and right barrier (red shaded region) where the regions either end at $z_-$ or start at $z_+$. 
    (b) Motion of the symmetrically kicked wave packet (initial width $\Delta z = \SI{3}{\text{\textmu}\meter}$, initial kick $p_0 = \pm 0.5 \times \sqrt{2 m V_b}$) prepared in the center of the matter-wave cavity for different gravitational accelerations.
    After a short period of time both wave packets are delocalized over the whole cavity where the white lines represent the position of the barriers. 
    The amplitude of the resulting oscillations decreases due to periodic outcoupling of the trapped atoms. 
    }
    \label{fig:ConfigAsymTunneling}
\end{figure}

To highlight the effect of gravitational fields on the matter-wave cavity, we remove the wave packet's propagation prior to the scattering and prepare a Gaussian wave packet in the center of the cavity, as shown in Figure~\ref{fig:ConfigAsymTunneling}(a). 
The width of the initial wave packet is chosen so that the overlap of the wave packet and the Gaussian barriers is negligible.
It is therefore completely confined inside the cavity.
Moreover, we create a superposition of wave packets with opposite momenta $\pm p_0$, e.g., via double Bragg diffraction~\cite{Giese2013, Giese2015, Hartmann2020}, modeled by displacing the wave packet in momentum representation. 

The fraction of atoms transmitted to the left and right differs because of the asymmetry of the matter-wave cavity in the gravitational field.
In analogy to the projector $\hat{P}_R$, we introduce the operator $\hat{P}_L^2 = \hat{P}_L$ that projects on the left space of the cavity, visualized in Fig.~\ref{fig:ConfigAsymTunneling}.
We define the projectors
\begin{equation}
    \hat{P}_\pm = \hat{P}_L \pm \hat{P}_R
\end{equation}
and the corresponding expectation values
\begin{equation}
    T_\pm = \bra{\psi_{sc}} \hat{P}_\pm \ket{\psi_{sc}} = \int_{-\infty}^{z_-} \dd{z} |\psi_{sc}(z)|^2 \pm \int_{z_+}^{\infty} \dd{z} |\psi_{sc}(z)|^2.
\end{equation}
where $z_-$ and $z_+$ are the position of the left and right barriers, respectively.
The total transmission $T_+$ approaches unity for times $t \to \infty$ independently of gravity, contrarily the asymmetric transmission $T_-$ depends on the gravitational acceleration. 
In addition, the variance of the asymmetric transmission $\Delta T_-^2 = T_+ - T_-^2$ depends on $T_-$ in a quadratic manner and thus the effect of gravity is enhanced, while the behavior $T_+ \to 1$ reduces it. 

\begin{figure}[h!]
	\centering
    \includegraphics{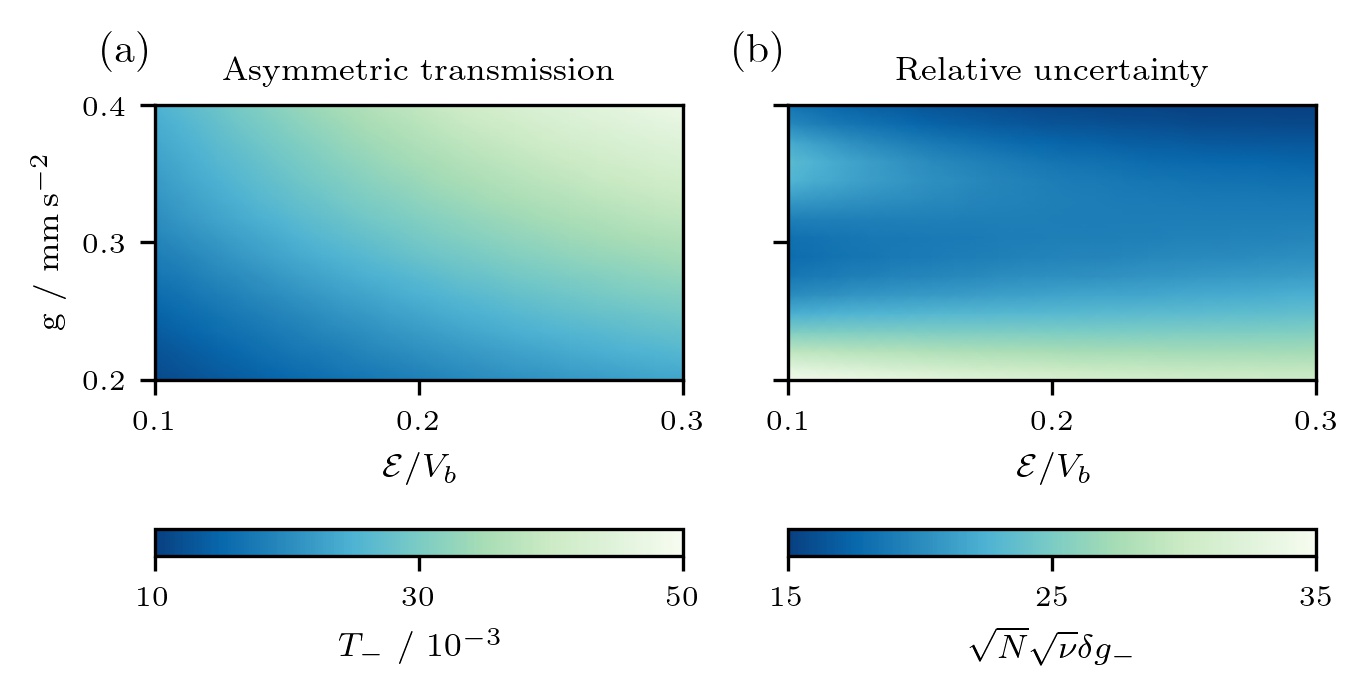}
    \caption{%
    Asymmetric transmission (a) and relative uncertainty (b) of a wave packet starting in the center of the gravitationally distorted matter-wave cavity ($\sigma_b = \SI{1}{\text{\textmu}\meter}$, $V_b=\SI{1.42e-25}{\joule}$, $d=\SI{15}{\text{\textmu}\meter}$).
    The wave packet (initial width $\Delta z = \SI{3}{\text{\textmu}\meter}$) experiences a double Bragg pulse, resulting in a superposition of two wave packets with opposite momenta $\pm p_0$ and kinetic energies $\mathcal{E} = p_0^2 / (2 m)$.
    The asymmetric transmission corresponds to the difference between the transmission through the left and right barrier of the matter-wave cavity.
    (a) No resonances are observed in the asymmetric transmission because of the large initial momentum width of the (localized) wave packet.  
    (b) The relative uncertainty associated with the asymmetric transmission contains a local maximum for small momentum kicks and shows the best sensitivity for the largest initial momentum and largest acceleration. 
    While $\delta g_-$ denotes the relative uncertainty for an experiment with $N$ particles and $\nu$ repetitions, we plot the quantity $\sqrt{N}\sqrt{\nu}\delta g_-$ which is the single-particle uncertainty without repetitions, assuming shot-noise limited measurements with non-interacting particles. 
    }
    \label{fig:AsymTransSens}
\end{figure}

Figure~\ref{fig:AsymTransSens} shows the asymmetric transmission $T_-$ and considers momentum transfers corresponding to kinetic energies larger than the energy $E_{r,2}$ associated with the third resonance of the matter-wave cavity.
Therefore, we set the end of the simulation to twice the lifetime $\Gamma_2^{-1}$ of the third resonance and thus only a negligible fraction of atoms are left inside the cavity.
Figure~\ref{fig:ConfigAsymTunneling}(b) shows a motion that resembles quantum carpets~\cite{Kaplan2000, Friesch_2000, Berry_2001}.
The initial wave packets are periodically reflected by the barriers leading to a standing wave inside the cavity whose amplitude decreases due to periodic outcoupling of the trapped atoms.
Moreover, the structure of the transmission spectrum washes out since the initial momentum width of the wave packet is larger than the distance between two resonances leading to an overlap of multiple resonances.
Consequently, we observe no peaks in the asymmetric transmission and obtain the largest asymmetries for the largest momentum transfer and largest acceleration, as shown in Fig.~\ref{fig:AsymTransSens}.

In analogy to the previous configuration, we consider for a perfectly centered wave packet with $z_0=0$ in analogy to Eq.~\eqref{eq:RelUncR} the relative uncertainty
\begin{equation}
    \delta g_- = \frac{\Delta g_-}{\sqrt{N} \sqrt{\nu} g} = \frac{\sqrt{T_+ - T_-^2}}{\sqrt{N} \sqrt{\nu} g \left| \partial g T_- \right|} 
\end{equation}
associated with the asymmetric transmission, shown in Fig.~\ref{fig:AsymTransSens} and observe the best sensitivity for the smallest initial momentum.
In addition, the structure of the transmission spectrum washes out and we obtain relative uncertainties one order of magnitude larger than preparing the wave packet outside the cavity.

So far we have assumed that the wave packet is initially centered inside the matter-wave cavity, consequently the contribution $| m g z_0 \partial_\mathcal{E} T_- |$ to the relative uncertainty vanishes for $z_0=0$.
However, in an actual experiment the centering will not be perfect.
If we assume that an initial displacement is smaller than 2\,\textmu m for our set of parameters, the contribution $| m g z_0 \partial_\mathcal{E} T_- |$ is much smaller than $ | g \partial_g T_- |$ and can be be neglected. 
In addition, the displacement influences wave packet effects induced by the nonlinear potential and subsequently affects the asymmetric transmission. 
For small momentum kicks such wave packet deformations are more prominent since they arise from different slopes of the Gaussian barriers in a linear potential.
A low velocity leads to a longer interaction time with these barriers.
However, this effect does not significantly change the sensitivity of the sensor and is suppressed for an increasing momentum transfer.

\section{Discussion}
\label{sec:Discussion}

We have proposed two setups that employ quantum tunneling in gravimetric applications. 
Since the cavity acts as a monochromator, the mean momentum and the momentum width at the time of scattering are crucial. 
Gravity influences the propagation of the wave packet prior to scattering and by that the momentum distribution at the time of interaction.
In addition, the slopes of the gravitationally distorted barriers affect the wave packet's width.
To remove the effect from propagation prior to the interaction, we have prepared the wave packet inside the matter-wave cavity and considered the asymmetric transmission as a measure for gravity. 
As a consequence, no optical counterpart exists.
Furthermore, the device has a relative uncertainty one order of magnitude larger than the one obtained from the monochromator setup, but also has a much lower susceptibility to a variation of the initial positions.

In addition to the momentum width of the wave packet, further effects limit the sensitivity of the gravimeter, including laser intensity fluctuations of the potential, different barrier heights, atom loss inside the gravimeter, heating, and nonlinear interactions of the atomic cloud. 
In particular, tight transverse confinement of the wave packet in the waveguide can give rise to additional contributions to the longitudinal motion depending on the scattering length as well as the length scales in longitudinal and transverse direction of the waveguide.
In addition, large dwell times and slow tunneling of a wave packet prepared inside the matter-wave cavity leads to low spatial densities of the tunneled wave packet and therefore limits the signal-to-noise ratio achievable at detection.
Moreover, imperfect preparation of the wave packet results in an uncertainty in the initial position and momentum.
In turn this leads to perturbations of the transmission and the corresponding sensitivity. 
This effect arises even if the wave packet is prepared inside the cavity.

To conclude, we have performed preliminary studies and have shown the feasibility of using matter-wave FPIs for accelerometry, using realistic cavities including gravitational distortions and the exact propagation of wave packets.
We have identified different effects that cause a susceptibility to gravity and laid the groundwork for quantum-technology based inertial sensors of this type.

\section*{Availability of data and materials}

All numerical code, used parameters and data necessary to reproduce the study are available from the authors on reasonable request.

\section*{Competing interests}
The authors declare that they have no competing interests.

\section*{Funding}
The QUANTUS and INTENTAS projects are supported by the German Space Agency at the German Aerospace Center (Deutsche Raumfahrtagentur im Deutschen Zentrum f\"ur Luft- und Raumfahrt, DLR) with funds provided by the Federal Ministry for Economic Affairs and Climate Action (Bundesministerium f\"ur Wirtschaft und Klimaschutz, BMWK) due to an enactment of the German Bundestag under Grant Nos. 50WM1956 (QUANTUS V), 50WM2250D-2250E (QUANTUS+), as well as 50WM2177-2178 (INTENTAS).
EG thanks the German Research Foundation (Deutsche Forschungsgemeinschaft, DFG) for a Mercator Fellowship within CRC 1227 (DQ-mat).
WPS is grateful to Texas A\& M University for a Faculty Fellowship at the Hagler Institute for Advanced Study
at Texas A\& M University and to Texas A\& M AgriLife for
the support of this work.
JRW is supported by the National Aeronautics and Space Administration through a contract with the Jet Propulsion Laboratory, California Institute of Technology.
Open Access funding enabled and organized by Projekt DEAL.

\section*{Authors' contributions}
\label{sec:author_contributions}
PS performed the main parts of the study, wrote the numerical code and ran the simulations.
The analysis and interpretation of data was performed together with AF and EG. 
PS also took a lead role in writing the manuscript together with AF and EG.
They also performed the conception and design of the study, while the idea of tunneling resonances as inertial sensor was developed by WPS and JRW, who also contributed to scientific discussions throughout the project. 
JRW provided input on experimental constraints and possible realizations.
EG and AF supervised the study. 

\section*{Acknowledgments}
\label{sec:acknowledgments}
We are grateful to M. O. Scully and T. Hoang for extensive discussions and their input in the preceding SURP project (JPL Task \# 01STSP/SP.20.0014.032) which lead to the initiation of the study presented in this article.
We also thank M. A. Efremov, E. P. Glasbrenner, D. Schlippert, A. Wolf, as well as the QUANTUS and INTENTAS teams for fruitful and interesting discussions.


\bibliographystyle{bmc-mathphys} 
\bibliography{bmc_article}      



\end{document}